\documentclass[%
twocolumn,reprint,
 amsmath,amssymb,
 aps,
]{revtex4-2}

\usepackage[utf8]{inputenc}
\usepackage{amsmath}
\usepackage{amssymb}
\usepackage{natbib}
\usepackage{graphicx} 
\usepackage{xcolor}
\usepackage{fullpage}

\begin{document}

\title{Stochastic Pairwise Preference Convergence in Bayesian Agents}
\date{May 2022}

\author{Jordan Kemp$^{1}$}
\thanks{These two authors contributed equally}
\author{Max-Olivier Hongler$^{2}$}
\thanks{These two authors contributed equally}
\author{Olivier Gallay$^3$}

 \affiliation{%
 $^1$Department of Physics, University of Chicago, Chicago, Illinois 60637, USA}%
\affiliation{%
 $^2$STI, Ecole Polytechnique F\'ed\'erale de Lausanne (EPFL), Lausanne 1015, Switzerland}%
\affiliation{%
 $^3$D\'epartement des Opérations, Universit\'e de, Lausanne 1015, Switzerland}%

\date{\today}

\begin{abstract}
    Beliefs inform the behavior of forward-thinking agents in complex environments.
    Recently, sequential Bayesian inference has emerged as a mechanism to study belief formation among agents adapting to dynamical conditions.
    However, we lack critical theory to explain how preferences evolve in cases of simple agent interactions.
    In this paper, we derive a Gaussian, pairwise agent interaction model to study how preferences converge when driven by observation of each other's behaviors.
    We show that the dynamics of convergence resemble an Ornstein-Uhlenbeck process, a common model in nonequilibrium stochastic dynamics.
    Using standard analytical and computational techniques,
    we find that the hyperprior magnitudes, representing the learning time, determine the convergence value and the asymptotic entropy of the preferences across pairs of agents. 
    We also show that the dynamical variance in preferences is characterized by a relaxation time $t^\star$, and compute its asymptotic upper bound.
    This formulation enhances the existing toolkit for modeling stochastic, interactive agents by formalizing leading theories in learning theory, and builds towards more comprehensive models of open problems in principal-agent and market theory.
\end{abstract}

\maketitle

\section{Introduction}

Belief formation is essential for studying behavior in the social and cognitive sciences.
In noisy environments, empirical beliefs are formed through observation \cite{seitz2020belief} and probabilistically predict future states to optimize energy costs \cite{friston2010free}.
Belief dynamics are critical for modeling how agents interact strategically in varying socio-economic contexts (through games), navigate uncertainty, and make decisions under imperfect information. 
However, there remain questions about how beliefs evolve in complex social environments such as networks \cite{dalege2023networks} and markets \cite{gjerstad1998price}, where the fluctuating beliefs (or perception of others') of asset values subject markets to intense volatility \cite{kurz2001endogenous} and divergent valuations \cite{farmer2020self}.

Several learning models have emerged to explain the formation of beliefs in stochastic multi-agent games \cite{shapley1953stochastic}, including frequentist and regression approaches \cite{fudenberg2016whither,evans2009learning}.
Reinforcement learning (RL) models are widely used and have intuitive descriptions \cite{bucsoniu2010multi,raileanu2018modeling}, but they do not produce closed-form solutions to dynamics of agent preferences \cite{sutton2018reinforcement}, hampering the search for generalizable results. 
These models are generally outperformed by learning frameworks based on Bayesian inference (BI) \cite{acuna2008bayesian,gershman2019believing,costa2015reversal}, where agents process information to inform history-dependent, optimally predictive, and (in some cases) analytically tractable models of their environment.
BI has thus become foundational in human cognition \cite{behrens2007learning,lee2014bayesian,gopnik2017changes, korb2010bayesian}, and in studying adaptive agent behavior in models of wealth and inequality \cite{kemp2023learning,bettencourt2019towards}, social dynamics \cite{pallavicini2021polarization}, and coordinated action \cite{kemp2023information,wu2021too}.
Additionally, Bayesian reversal learning has emerged as a more efficient alternative to RL in more realistic, non-stationary environments \cite{izquierdo2017neural} where discerning signal dynamics from noise is difficult \cite{d2011human,eckstein2022reinforcement}.

Solutions to closed-form belief  dynamics in stationary environments have contributed to a growing literature \cite{acuna2008bayesian,milani2007expectations,kemp2023learning}. 
However, they are not suitable for studying convergence in interacting models where signals are dynamic \cite{khan2014scalable,ignatenko2021preference}.
Studying pairwise dynamics in Gaussian models, for which analytical descriptions of distribution parameters exist \cite{murphy2007conjugate}, closes this theoretical gap while opening the door towards characterizing  emergent population preference dynamics \cite{farmer2020self,dalege2018attitudinal}.
This can be accomplished using established methods in non-equilibrium statistical physics, where the relationship between Bayesian inference and Ornstein-Uhlenbeck processes as noisy, mean-reverting processes with memory is well explored \cite{roberts2004bayesian,oravecz2016bayesian,singh2018fast}.
In the case of sequential Bayesian estimation, this analysis can used to study how convergence time relates to behavioral properties.  

In this paper, we propose a model for the statistical dynamics of two agents' preferences under Bayesian adaptation to  another's behaviors.
By treating behaviors as a Gaussian-distributed quantity, we can study the dynamics of preferences through the coupled Markov dynamics of its first-order moments.
We first show that in the absence of noise, the asymptotic preferences of the agents  converge to one another both to a relative value and on a timescale set by the relative strength of their priors.
Later, we introduce noise and show how the dynamics resemble an Ornstein-Uhlenbeck process with time-rescaling noise.
Using the Fokker-Planck equation (FPE), we then show that the preferences converge to a stationary distribution with a width set by the uncertainty in their behavior, and with dynamics governed by a relaxation time, $t^\star$. 
We conclude by discussing how convergence can be broken by introducing unpredictable behavioral shocks, and the model's implication for studying belief formation in principal-agent problems.

\section{Bayesian Preference Dynamics}

Consider agents  $A$ and $B$, who at time-step $i$ exhibit a statistically distributed, real-valued behavior $x_i\in X$ and $y_i\in Y$.
We denote the normalized  distribution of their decisions $P_i(X|\theta_A)$ and $P_i(Y|\theta_B)$, parameterized by behavioral parameters $\theta_A,\theta_B$. Consider that the agents can learn each other's behavior and are motivated to align their decisions (e.g., $x_i-y_i$ is minimized), but cannot directly coordinate their actions before observation.
While coordination can be accomplished by conditioning behavior on some shared signal \cite{kemp2023information}, this would not change the general dynamics and is excluded for brevity.

Each agent infers the other agent's preferences by observing their cumulative noisy behavior and adjusting their preferences to match.
By preferences, we mean the first moment of the distribution of behaviors that spans the agent's set of choices.
This particular setup is motivated by open questions in principal-agent problems, where agents must coordinate their behavior through adaptation.

History-dependent learning is accomplished optimally through BI \cite{kemp2023learning}.
As such, the distribution of agent $A$'s behaviors at $i=0$ forms a prior for their guess of $B's$, $P(X=x)\equiv P_0(\tilde Y=x)$, for approximated behavior $\tilde Y$ (and $\tilde X$ for $B$).
The distribution of decisions at later interactions is given by a posterior  $P_i(\tilde Y|\{y_i\})$, where the decision is conditioned on the history of $B$'s behavior 
\footnote{Through observation, $\tilde Y$ is conditioned on $\tilde X$, and agent $A$'s past behavior, filtered through $B$, influences their future behavior. $B$
is similarly conditioned by $A$'s.}. 
After $n$ steps, agent $A$'s posterior is given by (and $B$ by analogy) 

\begin{equation}
P_n(\tilde y|\{y_i\},\theta_A)=\bigg[\prod_{i=1}^n\frac{P(y_i|\tilde y_i)}{P(y_i)}\bigg]P(x|\theta_A),
\end{equation}

\noindent In sequential Bayesian inference, an agent's behavior at step $n$ follows a Markov process and is sampled from $P_{n-1}$.
This process is illustrated in Figure 1, where $\mathcal L$ denotes the likelihood of the evidence.

In this work, we assume the behaviors are instantaneously described by Gaussian distributions with gamma-distributed priors, 
$x\sim\mathcal{N}(\mu_x,\sigma_x|\boldsymbol\theta_x)$ and $y\sim\mathcal{N}( \mu_y,\sigma_y|\boldsymbol\theta_y)$, where $\boldsymbol\theta$ is the gamma prior vector.
The means,  $\mu_x$, $\mu_y$, describe the agents' preferences, whereas the fluctuation in true behavior is given by the Gaussian standard deviations  $\sigma_x$, $\sigma_y$. 

Bayesian inference on this choice of distribution results in preference dynamics that are linear \cite{murphy2007conjugate}.
Therefore, we first study the dynamics of the preference averages, then later consider how noisy behavior couples into the preference variances.
The following analysis gives a first-order approximation of the complete behavior ({\it vis a vis} the preferences) under Bayesian inference, whereas dynamics of higher order naturally come from higher-order moments and their couplings.
In this study, we will assume $\sigma_x=\sigma_y$ and leave the dynamics of the standard deviations under Bayesian inference for future work.

\begin{figure}
    \centering
    \includegraphics[width=.37\textwidth]{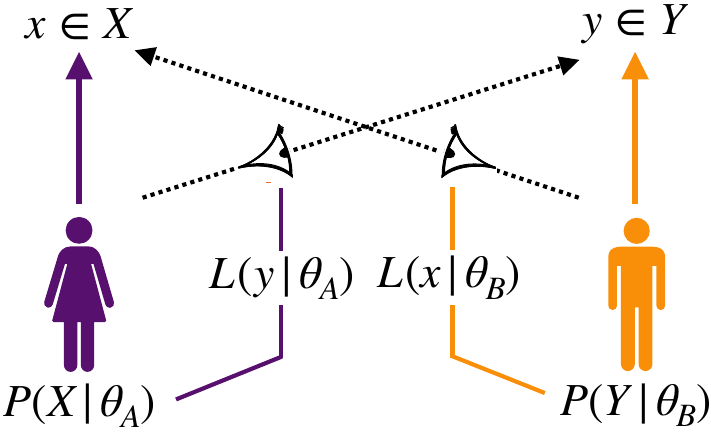}
    \caption{Diagram of the interaction model. Agents $A$  and $B$ sample behaviors $x\in X$ and $y\in Y$ from respective distributions. Agent $A$ updates their prior $\theta_A$ with the evidence $\mathcal L(y|\theta_a)$ from $B$'s behavior, and vice versa. }
    \label{fig:RecursMetric}
\end{figure}

\subsection{Deterministic  Dynamics}

First, we will study the dynamics of the preference parameter in the absence of noise.
The rule for updating the mean parameter of a Gaussian-Gamma model under Bayesian inference is described recursively after $n$ steps as (APP \ref{APP1}) \cite{murphy2007conjugate}

\begin{equation}\label{seconddef}
\begin{split}\nonumber
    \mu_x^n&=\frac{\mu_x^{n-1}\big(\tfrac{n-1}\omega+\alpha\big)+\mu_y^{n-1}}{\tfrac n\omega+\alpha},
    \hspace{.5cm}
    \mu_x^1=\frac{\alpha x_0+y_0}{1+\alpha},
\end{split}
\end{equation}

\noindent where $\omega=n/t$ is the interaction rate.
In the continuous limit  $\omega\rightarrow\infty$, $\mu_x$ and $\mu_y$ become coupled by the linear differential equations

\begin{equation}\label{fulldeterministic}
\frac{\partial \mu_x}{\partial t}=\frac{\mu_y(t)-\mu_x(t)}{t+\alpha},
\hspace{.5cm}
\frac{\partial \mu_y}{\partial t}=\frac{\mu_x(t)-\mu_y(t)}{t+\beta},
\end{equation}

\noindent where the hyperprior magnitudes $\alpha$, $\beta$, denoted the learning times, measure how resilient the preferences are to new evidence and have units $t$, and $x_0$, $y_0$ are the initial preferences.
These equations say that the dynamics of the preference parameters $\mu_x$, $\mu_y$  decrease as the quantities converge in time.
We demonstrate this by constructing the ODE for the difference measure $\Delta(t)=\mu_x(t)-\mu_y(t)$ (correspondingly $\Sigma(t)=\mu_x(t)+\mu_y(t)$), with solution (APP \ref{determino})

\begin{equation}\label{deltalimit}
\Delta(t)=\frac{\Delta_0\alpha\beta}{R(t)},
\end{equation}

\noindent where $\Delta_0=x_0-y_0$, and $R(t)=(\alpha+t)(\beta+t)$ is the time rescaling coefficient. 
This shows intuitively that the agents' preferences converge with power law $-2$ in time that increases symmetrically as $\alpha,\beta\rightarrow\infty$, and agent learning times increase.

With intuition for the coupled system established, we can now study the dynamics of the full system. 
There exist two solutions to Eqn. $\ref{fulldeterministic}$ given by the equality of the learning times.
First, when $\alpha=\beta$, the dynamics have the asymptotically symmetric solution $f(x_0,y_0,t)=\mu_x(t)$ and $f(y_0,x_0,t)=\mu_y(t)$, where $f$ is defined as

\begin{equation}
    f(x_0,y_0,t)=\frac{2\alpha^2x_0+(2\alpha t+t^2)(x_0+y_0)}{2(\alpha+t)^2}.
\end{equation}

\noindent It follows that $\displaystyle \lim_{t \rightarrow \infty}f=(x_0+y_0)/2$, and both agents' preferences converge to the average of their initial preferences asymptotically, at times $t\gg2\alpha$. 

In the case $\alpha\neq\beta$, the solution for $\mu_x$ is given by

\begin{equation}\label{dyndeteq}
\begin{split}
    \mu_x(t)=&
    \frac{\alpha x_0}{\alpha+t}
    +\frac{\alpha\beta(x_0-y_0)}{(\alpha-\beta)^2}K(t)+\frac{t(\alpha x_0-
    \beta y_0)}
    {(\alpha-\beta)(\alpha+t)}
\end{split}
\end{equation}

\noindent where $K(t)=\ln\big[\tfrac{\alpha\beta+\beta t}{\alpha\beta+\alpha t}\big]$ is a dynamical value with $\displaystyle \lim_{t \rightarrow \infty}K(t)=\ln[\beta/a]\equiv k_f$.
As we would expect, $\mu_x(0)=x_0$ and at long times, $\mu_x(t)\rightarrow s$, where $s$ is the weighted average between the initial values,

\begin{equation*}
s\equiv\frac{x_0\big[\alpha/\beta+k_f-1\big]+
    y_0\big[\beta/\alpha-k_f-1\big]}{(\alpha-\beta)^2/\alpha\beta}.
\end{equation*}

\noindent The solution for $\mu_y(t)$ is given in the appendix, with $\mu_y(t)\rightarrow s$ asymptotically. 
These results are demonstrated at the top of Fig. \ref{fig:fig2} for various learning times, with $y_0=5$ and $x_0=1$.
In matrix form, these dynamics are given by $\mathbb M[x_0,y_0]\equiv[x(t),y(t)]$, where the drift matrix is

  \begin{equation}\label{matdefmt}
  \begin{split}
  \mathbb M=\frac{\alpha\beta}{\alpha-\beta}\nonumber
  \begin{pmatrix}
       M_2(t)-M_1(0) & M_2(0)-M_2(t) \\
       M_1(t)-M_1(0) & M_2(0)-M_1(t),
  \end{pmatrix}\\
  M_1(t)=K(t)-\tfrac{1}{(t+\alpha)},\hspace{.2cm}M_2(t)=K(t)-\tfrac{1}{(t+\beta)}.
  \end{split}
    \end{equation}
This invertible matrix has a nonzero determinant $\det[\mathbb M(t)]=\alpha\beta/R(t)$.
As we will see, this gives the constant of motion for constructing exact solutions for the dynamics of the system with noise  \cite{chandrasekhar1943stochastic}.

\subsubsection{Asymptotic  preference behavior}

Conveniently, the asymptotic preference value can be expressed independently of the initial condition, allowing us to compute the relative shift in preferences as a function of learning times. 
Consider the initial parameter difference $\Delta_0$, and the difference in asymptotic value from the initial parameter $\delta_x=s-x_0$.
The fractional change in $X$, is given by $f_x=1-{\delta_x}/{\Delta_0}.$ 
This expresses how far $X$ has drifted from $x_0$ in the direction of $y_0$, and is useful for measuring the change in preferences of an agent represented by $X$ (and $Y$ by analogy).
It is given by

\begin{equation}\label{asymptotedetpref}
\begin{split}
    f_x&=1-
    \alpha\beta\frac{1-\alpha/\beta-\ln(\beta/\alpha)}{(\alpha-\beta)^2},\hspace{.2cm}f_y=1-f_x.
\end{split}
\end{equation}

\begin{figure}
    \centering
    \includegraphics[width=.44\textwidth]{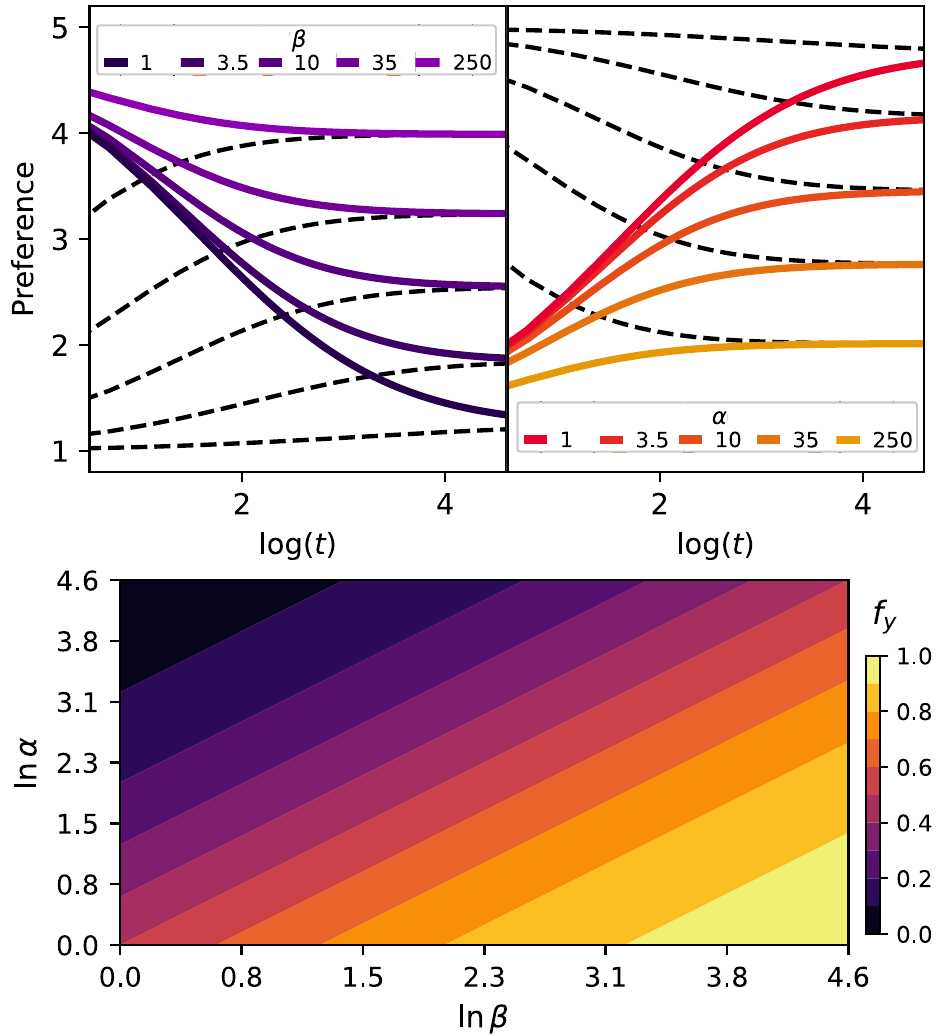}
    \caption{Dynamical and asymptotic behavior of the noiseless model in Eqn. \ref{fulldeterministic}. \textit{Top}: Convergence values computed from Eqn. \ref{dyndeteq} for variable $\beta$ (left) and variable $\alpha$ (right) for constant reference agent ($\alpha,\beta=5$), represented by dashed lines. \textit{Bottom:} Asymptotic fractional drift computed from \ref{asymptotedetpref} on a logarithmic parameter scale.}
    \label{fig:fig2}
\end{figure}

\noindent These fractional limits are demonstrated in the bottom of Fig. \ref{fig:fig2} over various learning times.

So far, we have explored the dynamics of this model without noise, and have shown that both preference parameters converge to a value set by the relative magnitude of the learning times.
We have shown that the deterministic dynamics are isomorphic and that we glean useful information about the relative change in preference between the agents without knowledge of the initial conditions.
These results establish intuition for how, on average, agent characteristics determine the convergence process.
In the following section, we will introduce noise to the inference process, and demonstrate a procedure for constructing exact solutions using the linear and isomorphic properties of the dynamics.
While this procedure results in lengthy analytical solutions that are not explored, we will demonstrate some key insights from the coupled dynamics, $\Delta(t),\Sigma(t)$.

\subsection{Full Dynamics under Noisy Sampling }

We introduce noise by rewriting Eqn. \ref{fulldeterministic} as the stochastic differential equations on quantities $X_t,Y_t$,

\begin{equation*}
dX_t=\tfrac{\sigma_y}{t+\alpha}dW_{2,t}-\tfrac{\Delta_t}{t+\alpha} dt,
\hspace{.2cm}
dY_t=\tfrac{\sigma_x}{t+\beta}dW_{1,t}+\tfrac{\Delta_t}{t+\beta}dt,
\end{equation*}

\noindent with boundary conditions $X_0=x_0,Y_0=y_0$.
We have introduced white Gaussian noise (WGN) processes, $dW_{1,t}$, $dW_{2,t}$ with magnitudes $\sigma_x$, $\sigma_y$ that describe i.i.d fluctuations in agent behavior.
Recalling previously that the asymptotic preferences depend on the initial conditions, we note that while the dynamics of the SDEs are Markovian, they cannot be  ergodic.
The dynamics of both preferences behave like Ornstein-Uhlenbeck (OU) processes, as the magnitude of the attractive drifts increases with the magnitude of the difference. 
However, this OU process is time inhomogeneous, as the magnitude of all dynamics decay with a power law in time.
We interpret these dynamics in terms of the underlying  Bayesian inference process.
The rate of parameter convergence slows as the agents converge in parameter value, and the effect of each interaction decreases in time as the agent weighs cumulatively larger sums of evidence. 
At long times, when preferences have nearly converged and have accumulated lengthy histories, small fluctuations dominate the dynamics.

To explore the statistics of the two-dimensional process, we define the bivariate transition probability distribution (TPD) as $P(x,y,t|x_0,y_0)$.
The evolution for this distribution is given by the FPE,
$\partial_tP(x,y,t|x_0,y_0)=\mathcal F\big[P\big]$, where $\mathcal{F}[\cdot]$ is defined 

\begin{equation}
\begin{split}
   \mathcal{F}[\cdot]=&\partial_x\big(\tfrac{y-x}{t+\alpha}[\cdot]\big)+\partial_y\big(\tfrac{x-y}{t+\beta}[\cdot]\big)\\
   &+\tfrac{\sigma_y^2}{2(t+\alpha)^2}\partial_{xx}[\cdot]+\tfrac{\sigma_x^2}{2(t+\beta)^2}\partial_{yy}[\cdot].
\end{split}
\end{equation}

\noindent One can marginalize the distribution for $x$, $P_M(x,t|x_0)=\int_{\mathbb R}P(x,y,t|x_0,y_0)dy$, and by analogy, $y$. 
To solve these equations exactly, we transform the set of equations into the frame of constant motion, defined by $\mathbb M(t)$, in which the process is purely diffusive and described by a Gaussian.
In this frame, solutions for the dynamics of $P_M(x,t)$ and $P_M(y,t)$ are exactly solvable \cite{chandrasekhar1943stochastic}.
However, this procedure does not lead to concise results and is detailed only in the appendix.

As in the deterministic case, we glean tractable insights into the dynamics by solving the FPE for the coupled system, $X_t\rightarrow\Delta_t=X_t-Y_t$, $Y_t\rightarrow\Sigma_t=X_t+Y_t$.
In the following section, we will use an exact solution of the FPE to show how the mean and variance of the TPD of $\Delta_t$ converges to zero, encoding the system's entropy into $\Sigma_t$. 
We will conclude this work by approximating an upper bound for the asymptotically stationary variance of $\Sigma(t)$.

\subsection{Solutions of the FPE for Coupled Dynamics}

In terms of the original model parameters, the new SDEs are

\begin{equation}\label{coupledSDE}
    \begin{split}
        d\Delta_t&=\tfrac{\sqrt{\sigma^2_y(t+\beta)+\sigma^2_x(t+\alpha)}}{R(t)}dW_{1,t}^\prime
        -\tfrac{2t+\alpha+\beta}{R(t)}\Delta_tdt\\
        d\Sigma_t&=\tfrac{\sqrt{\sigma^2_y(t+\beta)+\sigma^2_x(t+\alpha)}}{R(t)}dW_{2,t}^\prime+
        \tfrac{\alpha-\beta}{R(t)}\Delta_tdt,
    \end{split}
\end{equation}

\noindent where the  $dW^\prime$ terms are now correlated White Gaussian noise processes. 
Again, we see that the difference equation behaves like an OU process, where drift is set by the difference in preferences, with time-rescaling noise.
In this sense, $\Sigma_t$ does not couple into the dynamics of $\Delta_t$, permitting us to solve for the statistics of $\Delta_t$ first, then $\Sigma_t$.

\subsubsection{The Difference Equation}

In these coordinates, the statistics of $\Delta_t$ are fully described by the TPD $P_\Delta(z,t|z_0,0)=\textrm{Prob}\{z\leq\Delta_t\leq(z+dz)|z_0\}$, which solves the FPE
\begin{equation}\label{PDAL}
    \partial_tP_{\Delta}=\partial_z\big[\tfrac{2t+\alpha+\beta}{R(t)}zP_{\Delta}\big]+D(t)\partial_{zz}P_{\Delta},
\end{equation}

\noindent where the diffusivity $D(t)=\tfrac{\sigma_y^2(t+\beta)^2+\sigma^2_x(t+\alpha)}{2[R(t)]^2}$.
To solve this partial  differential equation, we seek the reference frame where the process becomes  purely diffusive. 
Consider the change of variables $z\mapsto z^\prime\equiv z\tfrac{
R(t)
}{\alpha\beta}$, and $t\mapsto\tau\equiv t$.
The differential operators transform as 
$\partial_z\mapsto\frac{R(t)}{\alpha\beta}\partial_{z^\prime},
\partial_t\mapsto\frac{2t+\alpha+\beta}{R(t)}z^\prime \partial_{z^\prime}+\partial_t$, where we used the equivalence $t=\tau\rightarrow\partial_t=\partial_\tau$, yielding
$\partial_tP_\Delta=\frac{2t+\alpha+\beta}{R(t)}P_\Delta+D(t)\tfrac{R(t)^2}{\alpha^2\beta^2}\partial_{z^\prime z^\prime}P_\Delta$ (APP \ref{CV1}).
Introducing the rescaling $P_\Delta\equiv R(t)Q_\Delta$, diffusion absorbs the  drift term and reduces the dynamics to time inhomogeneous diffusion
$\partial_tQ_\Delta =D^\prime(t) \partial_{z^\prime z^\prime}Q_\Delta$
where $D^\prime(t)=\frac{\sigma_x^2(t+\alpha)^2+\sigma_y^2(t+\beta)^2}{2\alpha^2\beta^2}$. To solve this equation, we introduce the time rescaling, $t\rightarrow s(t)=\int_0^tD(\xi)d\xi$, giving 

\begin{equation}
    s(t)=\frac{\sigma_x^2(t+\alpha)^3+\sigma_y^2(t+\beta)^3}{3\alpha^2\beta^2}-s_0\nonumber,
\end{equation}

\noindent where $s_0=\tfrac{\sigma_x^2\alpha}{\beta^2}+\tfrac{\sigma_y^2\beta}{\alpha^2}$.
Eqn. (\ref{PDAL}) has a Gaussian solution $Q_\Delta \sim \mathcal N\big(z_0^\prime,\sqrt{s(t)}\big)$.
This Gaussian transforms  back to the moving frame, $z^\prime\rightarrow z=z^\prime\alpha\beta/R(t)$ to get the full solution for $P_\Delta$

\begin{equation*}
P_\Delta(z,t|z_0,0)=\frac{R(t)}{\sqrt{2\pi\sigma_\Delta^2(t)}}\exp\bigg[-\frac{\big(z-z_0\tfrac{\alpha\beta}{R(t)}\big)^2}{2\sigma_\Delta^2(t)}\bigg],
\end{equation*}

\noindent where we have introduced the difference variance as 

\begin{equation}
    \sigma^{2}_\Delta(t)\equiv\frac{\sigma_x^2(t+\alpha)^3+\sigma_y^2(t+\beta)^3}{3R^2(t)}.
\end{equation}

\noindent This probability density  has a few key features.
First, $\displaystyle \lim_{t \rightarrow \infty}\langle \Delta_t\rangle=0$ and $\displaystyle \lim_{t \rightarrow \infty}\sigma_\Delta^2(t)=0$ so the distribution asymptotically converges to a delta function at $\Delta=0$.
By construction, $\sigma_{\Delta}(0)=0$  and  $\sigma^2_\Delta(t)\geq0$, meaning the variance evolves non-monotonously, and maximizes at a relaxation time $t^\star$.

\begin{figure}
    \centering

    \includegraphics[width=.40\textwidth]{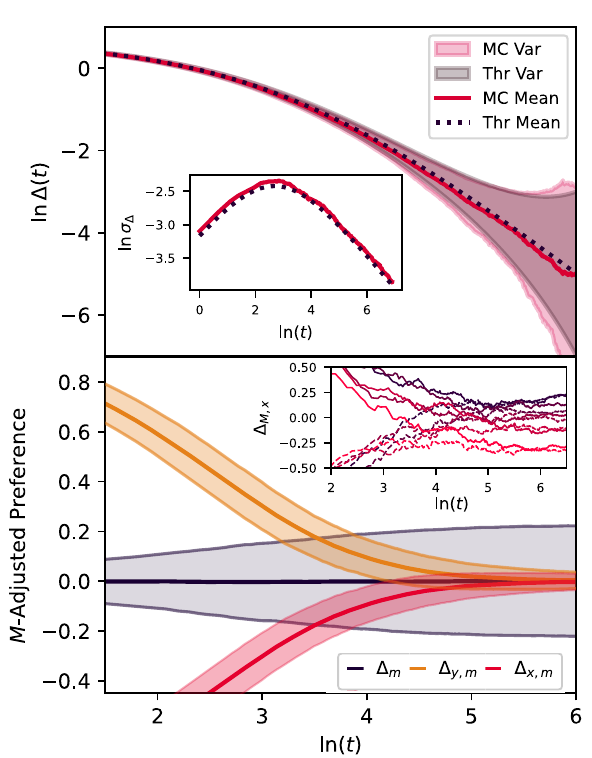}
    \caption{Coupled dynamics of stochastic agent preferences with $x_0=0,y_0=2$, $\alpha=\beta=25$ and 95\% CI shaded regions. \textit{Top:} MC Dynamics of $\Delta_t$ match theory. \textit{Inset:} The variance initially increases, reaches a maximum at $t^\star$, and then decreases. 
    \textit{Bottom:} The mean-adjusted dynamics ($\Delta_m=\Sigma/2$) is constant for this choice of parameters, with asymptotically constant noise. 
    The difference in agent parameters from the mean, $\Delta_m-y$, $\Delta_m-x$ converge to 0 with time vanishing noise. \textit{Inset}: Selected trajectories demonstrating different asymptotic values.
    }
    \label{fig:fig3}
\end{figure}

While cumbersome, these results are intuitive in terms of the underlying inference process. 
Agents' preferences converge to one another almost certainly in a time that increases with the magnitude of fluctuations but decreases in the strength of the agents' learning times. 
Therefore, the strength of attraction is asymptotically stronger than noise fluctuations.
These convergence dynamics are demonstrated by Monte Carlo (MC) simulations in Figure \ref{fig:fig3} with $N=1000$, $\alpha=\beta=15$, where we see that $\Delta\rightarrow0$ on a logarithmic scale in agreement with theory. 
We can define the observables $m(t)=\Sigma(t)/2$, $\Delta_m(t)=m(t)-m(0)$, where $\Delta_m$ measures how the mean of the dynamics change over time, and $\Delta_{y,m}=m(t)-y(t)$ demonstrates how $y$ differs from the mean. 
We see that $\Delta_{y,m}$ ($\Delta_{x,m}$) converges to zero with asymptotically vanishing noise, indicating that the preferences are converging to the mean almost certainly, while the entropy is increasingly expressed through the convergence value.

In the following section, we will use these insights to solve for the TPD of the $\Sigma_t$ process.

\begin{figure}
    \centering

    \includegraphics[width=.46\textwidth]{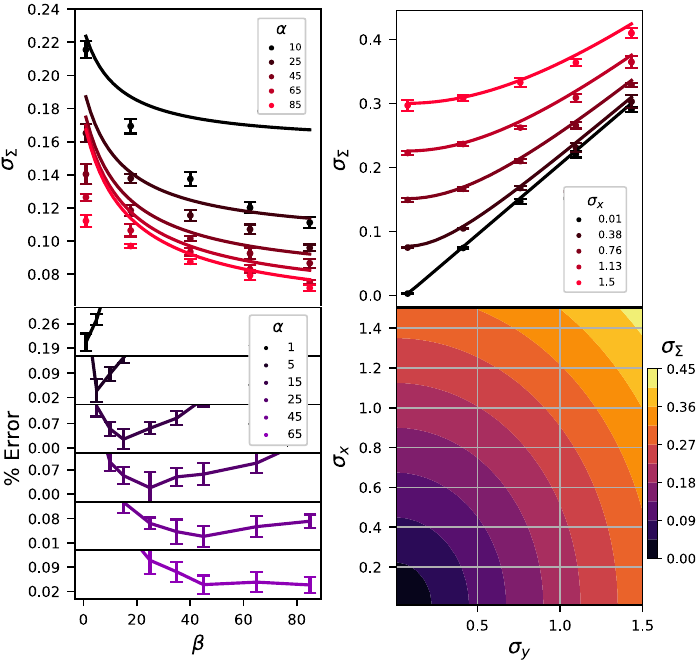}
    \caption{Asymptotic variance upper bound
    $\langle\Sigma^{2}_t\rangle$
    under various parameters
    \textit{Top:} $\langle\Sigma^{2}_t\rangle$ increases with agent noise with agreement between MC simulations and theory.
    \textit{Bottom:} Variance decreases with agent learning time. \textit{Left:} Variance upper bound  $\langle\Sigma^{2}_t\rangle$, diverges from empirical results as $\beta$ and $\alpha$ diverge. \textit{Right:} Deviation between the upper bound and MC experiments as a fraction of theoretical prediction.
    }
    \label{fig:fig4}
\end{figure}

\subsubsection{The Sum Equation}

The cumbersome  dynamics of $\Delta_t$  lead to an even more complex analytical description for $\Sigma_t$. However, we know that $\displaystyle \lim_{t \rightarrow \infty}\langle\Delta\rangle(t)=0$ and $\displaystyle \lim_{t \rightarrow \infty}\sigma_ \Delta(t)=0$.
It follows that the initially bi-variate diffusion process asymptotically collapses into a uni-variate pure diffusion centered at $\mu_{\Sigma,f}$. Hence, for $t \rightarrow \infty$, we shall approximately have (see APP \ref{SIGMATIME})

\begin{equation}\label{coupledSDE}
    \begin{split}
        d(\Sigma_t- \mu_{\Sigma,f})=d\Sigma_t&\approx \tfrac{\sqrt{\sigma^2_y(t+\beta)+\sigma^2_x(t+\alpha)}}{(t+\beta)(t+\alpha)}dW_{2,t}^\prime,\nonumber
    \end{split}
\end{equation}

\noindent where the constant $\mu_{\Sigma,f}$ is given in Eq.(\ref{LIMO}). The corresponding TPD  $P_\Sigma(z,t|\mu_{\Sigma,f})dz=\textrm{Prob}\{z\leq\Sigma_t\leq(z+dz)|\sigma_0\}$ solves the FPE:

\begin{equation*}\partial_tP_\Sigma=\tfrac{\sigma_x^2(t+\beta)^2+\sigma_y^2(t+\alpha)^2}{2R(t)}\partial_{zz}P_\Sigma.
\end{equation*}

\noindent By inspection,  $P_\Sigma$ is a Gaussian law  with  mean  $\mu_{\Sigma,f}$ and by an ad-hoc time re-scaling, (see Appendix \ref{SIGMATIME}),  we obtain the time-dependent variance as

\begin{equation}\label{noiseub}
\begin{array}{l}
    \langle\Sigma_t^{2}\rangle=\sigma_x^2\big[\tfrac{1}{\alpha}-\tfrac{1}{\alpha+t}\big]+\sigma_y^2\big[\tfrac{1}{\beta}-\tfrac{1}{\beta+t}\big], \\\\
    \end{array}
\end{equation}

\noindent We now see that the second moment (and hence the  variance)  converges in the limit $t\rightarrow\infty$ to a stationary value $\displaystyle \langle\Sigma_t^{2}\rangle=\tfrac{\sigma_x^2}{\alpha}+\tfrac{\sigma_y^2}{\beta}$. 
Similarly to the case of $\sigma^{2}_\Delta$, the variance of this distribution increases with the fluctuations in behavior and decreases with hyperprior strength. 
We note, though, that this is an upper-bound estimation for variance, as

\begin{equation*}
 \sigma^2_\Sigma(t)=\langle (\Sigma_t - \mu_{\Sigma,f} )^{2}\rangle = 
    \langle\Sigma^{2}_t\rangle - \mu_{\Sigma,f}^{2}\leq 
    \langle\Sigma_t^{2}\rangle.
\end{equation*}
\noindent Furthermore, from the convergence of $\Delta_t\rightarrow0$, we know preferences converge to $X_t=\displaystyle Y_t$, and the variances converge to $\langle X_t^2\rangle=\displaystyle \langle Y_t^2\rangle=\tfrac{\langle\Sigma^2_t\rangle}{2}$ as $t\rightarrow\infty$.


The asymptotic  preference variance demonstrated in Eqn. \ref{noiseub} shows that  more noisy agents who learn quickly will converge to more entropic states.
The relationships between these quantities and the variance are demonstrated by MC simulations in Fig. \ref{fig:fig4}.
We also see through simulations that the percent error in the upper bound estimate and the true variance are closest when $\beta=\alpha$, and increases as the learning times diverge.
However, this difference decreases as both learning times approach large values. 

\section{Discussion}

In this paper, we studied a simple model for pairwise belief formation in Bayesian agents who adapt to each other's behaviors.
We showed that preferences converge on a timescale and to a value given by the agents' relative learning times.
Using the Fokker-Planck equation we then explored the convergence characteristics of the Gaussian PDF for  preferences in the combined frame.
We showed that while agents' preferences invariably converge to one another, the relative value is noisy, is characterized by a relaxation time $t^\star$, and  is bounded above by a sum of the standard deviations of agent behaviors weighted by the learning times.

There remain several challenges concerning the full characterization of this system. 
First, deriving the full dynamics of $\Sigma_t$ would be useful for attaining better bounds on the asymptotic coupled behavior.
Second, solving the nonlinear dynamics for the covariance matrix of the inference process would give the full dynamics of the interaction, although it would likely require numerical treatment. 
Once we understand the full dynamics of this interaction, we can scale this model to include agents with multidimensional, co-varying preferences.
This builds towards a Bayesian analog of a Self-Other Model \cite{raileanu2018modeling}, wherein agents coordinate decisions by approximating the other agent's behavior and serve as a microfoundation for more robust statistical mechanical models of network belief formation \cite{dalege2018attitudinal,dalege2023networks}.  
Furthermore, we can extend this analysis by studying preference dynamics in agents that must balance learning each other's signals with some additional, external signals.
When only one agent observes an additional, stationary signal, it is natural that the agents' preferences would converge to a value biased by the external signal.
However, when one agent observes an additional, non-stationary signal, as a form of unpredictable shock, or both agents observe separate, stationary signals as a form of "reality check", their preferences are not guaranteed to converge \cite{farmer2020self}.
The existence of a phase transition would depend on whether the external signal alters their preferences on a timescale comparable to the relaxation time $t^\star$, and can be applied at scale to study many-body preference dynamics.

Although, this work can already be applied to open problems in the game theory literature. This formalism can be adapted to study how preferences evolve in principal-agent models where there remain questions of how the value and variance of asymptotic preferences behave as agents adapt to post-contract disagreements.
In cases where the agent serves as an information channel for the principal, 
models of information-driven resources dynamics \cite{kemp2023learning} can be used to study how the convergence rate affects agent resources, and how the entropy of convergence values affects the quality of information transmitted to the principal.
Generally, these results constitute a step towards more robust quantitative models of inter-agent and market interactions that incorporate findings from the cognition community. 

We thank Adam Kline, Andrew Stier, and Lu\'is Bettencourt for their discussions and comments on the manuscript.
This work is supported by the  Department of Physics at the University of Chicago, the National Science Foundation Graduate Research Fellowship (Grant No. DGE 1746045 to JTK), and the Swiss State Secretariat for Education,
Research and Innovation SERI (to JTK). 
\begin{widetext}
\appendix

\section{Defining the Deterministic ODE}\label{APP1}

The equation for the mean of the Gaussian Gamma for variable $x$ is given by 

\begin{equation}\nonumber
    \mu_x^n=\frac{\sum_{i=0}^ny_i+\mu_0\alpha}{n+1+\alpha} =\frac{y_{n-1} +\sum_{i=0}^{n-1}y_i+\mu_0\alpha}{n+1+\alpha}
\end{equation}

The sample $y_n$ corresponds to the mean of $P(y)$, $\mu_y$, plus some noise $\xi_n$.
In the deterministic case, $\xi_n=0$, and remaining two terms constitute $\mu_x^{n-1}(n+\alpha)$.
We can therefore redefine this quantity as Eqn. \ref{seconddef} from the main text

\begin{equation}
    \begin{split}
        \mu_x^n&=\frac{\mu_x^{n-1}\big[(n-1)/\omega+\alpha\big]+\mu_y^{n-1}}{\tfrac n\omega+\alpha},
    \end{split}
\end{equation}

\noindent where we have written $n\rightarrow \tfrac n\omega$ to enable a conversion to continuous time variables later on.
In the deterministic case, we define the difference operator

\begin{equation}
\begin{split}
    \Delta\mu_x=\mu_x^n-\mu_x^{n-1}&=\frac{\mu_x^{n-1}\big[(n-1)/\omega+\alpha\big]+\mu_y^{n-1}}{\tfrac n\omega+\alpha}
    -\frac{\mu_x^{n-1}(\tfrac n\omega+\alpha)}{\tfrac n\omega+\alpha}\\
    &=\frac{\mu_y^n-\mu_x^n}{\tfrac n\omega+\alpha}
\end{split}
\end{equation}

\noindent which as $\omega\rightarrow\infty$ converges to the  continuous time ODE used in the main text. Finally, Gaussian noise will be   linearly added to the deterministic  dynamics. 

\subsection{Deterministic evolution of preferences}\label{determino}

\noindent We start by solving the deterministic motion, namely the set of ODEs:

\begin{equation}
\label{BB1}
\left\{
\begin{array}{l}
\frac{d\mu_x(t) }{dt}= \frac{\mu_y(t)- \mu_x(t) }{t + \alpha}, \qquad \mu_x(0) = x_0 \\\\ 

\frac{d\mu_y(t)}{dt} = \frac{\mu_x(t)- \mu_y(t) }{t + \beta},  \qquad \mu_y(0) = y_0.
\end{array}
\right.
\end{equation}

\noindent  To process further, we introduce the new variables:

\begin{equation}
\label{SIGDEL}
 \mu_{\Delta}(t) = \left[ \mu_x(t)- \mu_y(t)\right]\qquad {\rm  and}  \qquad \mu_{\Sigma}(t) = \left[ \mu_x(t) + \mu_y(t)\right].
\end{equation}

\noindent In terms of the new variables, Eqn. (\ref{BB1}) reads:

\begin{equation}
\label{BB2}
\left\{
\begin{array}{l}
\left[ (t+ \alpha) (t+ \beta) \right] \frac{d\mu_{\Delta}(t)}{dt } = -  \left[2t + \alpha + \beta \right] \mu_{\Delta}(t),  \\\\ 
\left[ (t+ \alpha) (t+ \beta) \right] \frac{d\mu_{\Sigma}(t) }{dt}= \left[\alpha - \beta \right] \mu_{\Delta}(t) .
\end{array}
\right.
\end{equation}

\noindent From Eqn. (\ref{BB2}), we immediately have:

\begin{equation}
\label{BB3}
\left\{
\begin{array}{l}
 \frac{d \ln(\mu_{\Delta}(t) )}{dt} = - \frac{\left[2t + \alpha + \beta \right]} {\left[ (t+ \alpha) (t+ \beta) \right]  }  = -\frac{d \ln \left[ (t+ \alpha) (t+ \beta) \right] }{dt} \,\,\, \Rightarrow \,\,\,  \mu_{\Delta}(t) = \frac{\Delta_0 \alpha \beta }{ \left[ (t+ \alpha) (t+ \beta) \right]}: = \frac{\Delta_0 \alpha \beta }{R(t)},  \\\\ 
\left[ (t+ \alpha) (t+ \beta) \right] \frac{d\mu_{\Sigma}(t) }{dt}= \left[\alpha - \beta \right] \mu_{\Delta}(t)  \quad \Rightarrow \quad  \frac{d\mu_{\Sigma}(t)}{dt} = \frac{(\alpha- \beta) \alpha \beta \Delta_0}{ \left[ (t+ \alpha) (t+ \beta) \right] ^{2}}  = \Delta_0 \frac{\alpha \beta (\alpha - \beta)}{R^{2}(t)} 
\end{array}
\right.
\end{equation}

\noindent with:

\begin{equation}
\label{DEFINEX}
R(t) : =  \left[ (t+ \alpha) (t+ \beta) \right]
\end{equation}

  \noindent By direct calculation, one may verify   the couple of  identities:

\begin{equation}
\label{ID}
\begin{array}{l}
\int\frac{1}{R(t)}  dt =   \frac{1}{\alpha - \beta}\ln \left[ \frac{t+ \beta}{t + \alpha }\right] 
\\\\
\int\frac{1}{R^{2}(t)}  dt =- \frac{1}{(\alpha - \beta)^{2}}  \left\{  \frac{d\ln [R(t)]}{dt}  + 2 \int\frac{1}{R(t)}  dt  \right\} = 
- \frac{1}{(\alpha - \beta)^{2}}  \left\{  \frac{d \ln [R(t)]}{dt}  + \frac{2 }{\alpha - \beta} \ln \left[ \frac{t+ \beta}{t + \alpha }\right]  \right\} = \\\\

\qquad \qquad -\frac{1}{(\alpha - \beta)^{2}}  \left\{ \frac{2t + \alpha + \beta}{t^{2} + (\alpha + \beta) t + \alpha \beta}  + \frac{2 }{\alpha - \beta} \ln \left[ \frac{t+ \beta}{t + \alpha }\right]\right\} .
\end{array}
\end{equation}

\noindent 
Accordingly, from Eqns. (\ref{BB3}) and (\ref{ID}),  we obtain:
\begin{equation}
\label{DETEXGAMMA}
\mu_{\Sigma}(t) =  \int\frac{(\alpha -\beta)\alpha \beta\Delta_0}{R^{2}(t)}  dt =  -\alpha \beta \Delta_0  \left\{ 
\frac{2t + \alpha + \beta}{(\alpha - \beta) R(t)} + \frac{2}{(\alpha - \beta)^{2}}  \ln \left[\frac{t + \beta}{t + \alpha }\right]
\right\} + C
\end{equation}

\noindent where $C$ is an integration  constant to be determined by the initial condition. At time $t=0$, we have:

\begin{equation}
\label{SIGMAZERO}
C = \Sigma_{0} + \Delta_{0}\left\{ \frac{\alpha + \beta}{ (\alpha - \beta)}  +   \frac{2 \alpha \beta [\ln\beta - \ln \alpha]}{(\alpha - \beta)^{2}}\right\}.
\end{equation}

\noindent Hence we can write:

\begin{equation}
\label{SIG1}
\mu_{\Sigma}(t) = \Sigma_0 +\frac{ \Delta_0}{(\alpha - \beta)}  \left[\alpha+ \beta - \frac{(2t + \alpha + \beta) \alpha \beta }{(t + \alpha)(t+ \beta)}  + \frac{2\alpha \beta }{(\alpha - \beta)} \ln \left(\frac{\beta t + \alpha \beta}{\alpha t + \alpha \beta}  \right) \right] .
\end{equation}

\noindent Note  that we have:

\begin{equation}
\label{LIMO}
\left\{
\begin{array}{l}
\displaystyle \lim_{t \rightarrow 0^{+}} \mu_{\Sigma}(t) = \Sigma_0, \\\\
\displaystyle \lim_{t \rightarrow \infty} \mu_{\Sigma}(t):= \mu_{\Sigma, f }= \Sigma_0 + \Delta_0\left[  \frac{\alpha+ \beta}{\alpha -\beta}\right]= 2 \left[\frac{\alpha x_0 -  \beta y_0}{\alpha - \beta}\right]+ 2\Delta_0 \frac{\alpha \beta \ln[ \beta/\alpha]}{(\alpha- \beta)^{2}}, \hspace{.5cm}\alpha\neq\beta,\\\\
\displaystyle \lim_{t \rightarrow \infty} \mu_{\Sigma}(t):= \Sigma_0, \hspace{.5cm}\alpha=\beta,
 \end{array}
\right.
\end{equation}

\noindent In terms of $\mu_{x}(t)$ and $\mu_{y}(t)$, we have:

\begin{equation}
\label{SOLUT}
\left\{
\begin{array}{l}
\mu_x(t) =  \frac{\mu_{\Delta}(t) + \mu_{\Sigma}(t)}{2}=  \left[\frac{\alpha x_0-  \beta y_0}{\alpha - \beta}\right] +  (x_0- y_0)\left\{\frac{\alpha \beta}{\alpha -\beta}  \ln \left[ \frac{\alpha \beta + \beta t}{\alpha \beta + \alpha t}\right] -\frac{ \alpha \beta}{ (\alpha - \beta)(t+ \alpha) }  \right\} .
\\\\
\mu_y(t)\,=   \frac{\mu_{\Delta}(t) - \mu_{\Sigma}(t)}{2} =  \left[\frac{\alpha x_0 -  \beta y_0}{\alpha - \beta}\right] + (x_0- y_0)\left\{\frac{\alpha \beta}{\alpha -\beta}  \ln \left[ \frac{\alpha \beta + \beta t}{\alpha \beta + \alpha t}\right] -\frac{ \alpha \beta}{ (\alpha - \beta)(t+ \beta) }  \right\} 

\end{array}
\right.
\end{equation}

\noindent  which can  be summarised  as:

\begin{equation}
\label{MATROS}
\begin{pmatrix}
      \mu_x(t)  \\\\
     \mu_y(t) 
\end{pmatrix} = 
\mathbb{M}(t) \begin{pmatrix}
      x_0    \\\\
      y_0 
\end{pmatrix},
\end{equation}
\noindent where  the matrix $\mathbb{M}(t)$ reads as:

\begin{equation}
\label{MATDEFO}
\begin{array}{l}
\mathbb{M} = 
\begin{pmatrix}
 -M_1(0)+ M_2(t)    & 
   M_2(0) - M_2(t) \\\
       -M_1(0) + M_1(t) &     M_2(0) - M_1(t) 
       \end{pmatrix}, \\\\
       M_1(t) =   \frac{\alpha \beta}{\alpha -\beta}  \ln \left[ \frac{\alpha \beta + \beta t}{\alpha \beta + \alpha t}\right]  - \frac{\alpha \beta }{(\alpha - \beta) (t + \alpha)},\\\\

        M_2(t) =  \frac{\alpha \beta}{\alpha -\beta}  \ln \left[ \frac{\alpha \beta + \beta t}{\alpha \beta + \alpha t} \right]  - \frac{\alpha \beta }{(\alpha - \beta) (t + \beta)} .
       \end{array}
\end{equation}


\subsubsection*{Stationary regime}
\noindent  From Eqns. (\ref{MATROS}) and (\ref{MATDEFO}), one  immediately concludes that the final state $\left( \mu_x(\infty),  \mu_x(\infty) \right)$ is given by:

\begin{equation}
\label{MATROSINF}
\begin{pmatrix}
      \mu_x(\infty) \\\\
     \mu_y(\infty)
\end{pmatrix} = 
\begin{pmatrix}
  \frac{\alpha} {\alpha - \beta}  & 
       \frac{\beta}{\beta - \alpha} \\\\
        \frac{\alpha} {\alpha - \beta} &      \frac{\beta}{\beta - \alpha}\end{pmatrix}
\begin{pmatrix}
      x_0    \\\\
      y_0 
\end{pmatrix}:= \mathbb{M}_{\infty}\begin{pmatrix}
      x_0    \\\\
      y_0 
\end{pmatrix}
\end{equation}

\noindent  We observe that  $\left(\mu_x(\infty), \mu_y(\infty) \right)$ depends on the initial condition $(x_0, x_0)$.  We note that $\alpha = \beta$ is a singular situation. In addition observe also that for  $\alpha =0$,   we obtain $\mu_x(\infty) = \mu_y(\infty)= y_0$ and conversely  for $\beta =0$, we have  $\mu_x(\infty) = \mu_y(\infty)= x_0$ thus showing that in both of these  limiting  cases, the evolution affects a single variable.

\section{Solving the TPD using Liouville coordinates}\label{LIOUVILLE}

\noindent This Appendix  shows that an ad-hoc  time-dependent change of coordinates $(X_t, Y_t) \mapsto (U_t, V_t)$ transforms the nominal drifted process into a pure bi-variate diffusion for which  the analytical  probability density can be calculated.  First one observes:

\begin{equation}
\label{DERTERMIN}
{\rm Det} \left[  \mathbb{M}(t) \right] = \frac{\alpha \beta }{ R(t)} \neq 0 \, \,\,{\rm for} \,\,\, 0 <t< \infty,
\end{equation}

\noindent with $R(t)$ as defined in Eqn. (\ref{DEFINEX}). Hence  the inverse matrix ${\cal M}(t)$ exists an reads:

\begin{equation}
\label{MOINS}
\left\{
\begin{array}{l}
\mathbb{M}(t){\cal M}(t)= {\cal M} (t)\mathbb{M}(t)=\mathbb{I}{\rm d}, \\\\
 {\cal M} (t)=   \frac{R(t)} {\alpha \beta} \begin{pmatrix}
  M_2(0) - M_1(t)   &   -M_2(0) + M_2(t)   \\\\
  M_1(0) - M_1(t)    &  -M_1(0)+ M_2(t)  \end{pmatrix}:= \begin{pmatrix}
     m_{11}(t) &   m_{12}(t)  \\
       m_{21}(t)&   m_{22}(t)
\end{pmatrix}.
  \end{array}
  \right. 
\end{equation}

\noindent In particular using Eqn. (\ref{MATROS}),  we can write:

\begin{equation}
\label{ABB1}
\begin{pmatrix}
     x_0\\\\
     y_0
\end{pmatrix} =  {\cal M} (t)
\begin{pmatrix}
      \mu_x(t)   \\\\
      \mu_y(t)
\end{pmatrix},
\end{equation}

\noindent where the initial values  $(x_0, y_0)$ are constants of the motion. This suggests to  introduce the  time-dependent   change of coordinates (i.e. Liouville coordinates)  defined by:

\begin{equation}
\label{CCO}
\left\{
\begin{array}{l}

{\bf z}:=\begin{pmatrix}
      x    \\
      y 
\end{pmatrix}\mapsto {\bf w} := \begin{pmatrix}
      u    \\
      v  
\end{pmatrix}, \\\\
\begin{pmatrix}
     u     \\
      v 
\end{pmatrix} = 
{\cal M} (t)\begin{pmatrix}
      x   \\
      y  
\end{pmatrix} := \begin{pmatrix}
     m_{11}(t) & m_{12}(t)   \\
     m_{21} (t)&  m_{22}(t)
\end{pmatrix}\begin{pmatrix}
     x   \\
      y
\end{pmatrix}.

\end{array}
\right.
\end{equation}

\noindent In terms of the ${\bf w}$  coordinates the motion is a purely diffusive, (i.e. the drift components  in the Fokker-Planck equation cancel out) and  we have:

\begin{equation}
\label{OPERATORS}
\left\{
\begin{array}{l}
\partial_t [\cdot]   \mapsto  \partial_t  [\cdot]  + \frac{\partial_u}{\partial_t}  \partial_u[\cdot] +  \frac{\partial_v}{\partial_t} \partial_v  [\cdot],  \\\\
\begin{pmatrix}
      \partial_x  [\cdot]  \\\\
      \partial_y   [\cdot] 
\end{pmatrix} \mapsto {\cal M}^{\dagger} (t) \begin{pmatrix}
      \partial_u   [\cdot]   \\
      \partial_v  [\cdot] 
\end{pmatrix} , \\\\ 

\frac{1}{2}   \left[ \partial_{xx} +
      \partial_{yy}\right]  [\cdot] 
\mapsto  \triangle_{uv}  [\cdot] :=  \frac{1}{2} \begin{pmatrix}
      \partial_u  [\cdot], &    \partial_v [\cdot] 
    
\end{pmatrix}{\cal M} (t) {\cal M}^{\dagger} (t) \begin{pmatrix}
      \partial_u  [\cdot]    \\
      \partial_v  [\cdot] 
\end{pmatrix}, \\\\
\partial_t P = \triangle_{uv} [P], \\\\
P = \frac{1}{2 \pi \sqrt{{\rm Det}[ \Pi(t)]}} e^{- {\bf w}^{\dagger} \frac{\Pi(t) }{{\rm Det}(\Pi(t))}{\bf w} }.
\end{array}
\right.
\end{equation}

\noindent  Eqn. (\ref{OPERATORS}) describes the  TPD evolution of the  pure diffusion process:

\begin{equation}
\label{BIVAR}
\begin{pmatrix}
      dU_t    \\
      dV_t
\end{pmatrix} = {\cal M} (t)\begin{pmatrix}
      dB_{1,t}  \\
       dB_{2,t} 
\end{pmatrix}.
\end{equation}

\noindent  where  $dB_{1,t} $ and $dB_{2,t} $ are independent White Gaussian Noise (WGN) processes. The Fokker-Planck equation  in  Eqn.(\ref{OPERATORS}) describes the TPD of the bi-variates $(U_t, V_t)$ pure Gaussian  process and we have:

\begin{equation}
\label{PIPO}
\begin{array}{l}
\Pi(t) :=  
 \begin{pmatrix}
     \mathbb{E}\left\{V_t^{2}\right\} & \mathbb{E}\left\{V_t U_t\right\}   \\
     \mathbb{E}\left\{U_t V_t \right\}&  \mathbb{E}\left\{U_t^{2}\right\}  \end{pmatrix}  := \begin{pmatrix}
      \pi_{11}(t)&    \pi_{12}(t)\\
    \pi_{21} (t) &  \pi_{22}(t)
\end{pmatrix},
\end{array}\end{equation}

\begin{equation}
\label{LIST}
\left\{
\begin{array}{l}
\mathbb{E} \left\{ dU_t ; dU_{\tau}\right\} = \mathbb{E} \left\{\left[m_{11}(t) dB_{1,t} + m_{12 }(t)dB_{2,t}\right];  \left[ m_{11}(\tau) dB_{1,\tau} +  m_{12} (\tau)dB_{2,\tau} \right]\right\} = \\\\
\qquad  \qquad \qquad \qquad \qquad \qquad \qquad \left[m_{11}(t) m_{11}(\tau)  + m_{12 }(t) m_{12 }(\tau)\right]\delta(t- \tau), \\\\

\mathbb{E} \left\{ dU_t;dV_{\tau}\right\} =  \left[m_{11}(t) m_{12}(\tau)  + m_{22 }(t) m_{21}(\tau)\right]\delta(t- \tau), \\\\
\mathbb{E} \left\{ dV_t; dU_{\tau}\right\} =  \left[m_{11}(t) m_{21}(\tau)  + m_{22 }(t) m_{12}(\tau)\right]\delta(t- \tau),
\\\\
\mathbb{E} \left\{ dV_t; dV_{\tau}\right\} =  \left[m_{22}(t) m_{22}(\tau)  + m_{21 }(t) m_{21}(\tau)\right]\delta(t- \tau).
\end{array}
\right.
\end{equation}

\noindent Invoking Theorem 3.6 of  A. H. Jazwinsky \footnote{A. H. Jazwinsky, {\it Stochastic Processes and Filtering Theory}. Acad. Press. (1970). See the entry 3.67).}, 
Eqn. (\ref{LIST}) leads to:

\begin{equation}
\label{LIST1}
\left\{
\begin{array}{l}
\mathbb{E} \left\{ U^{2}_t\right\} = \int_{0}^{t} ds \int_{0}^{s} d\tau\left[m_{11}(s) m_{11}(\tau)  + m_{12 }(s) m_{12 }(\tau)\right]\delta(s- \tau) = \\\\

\qquad \qquad \qquad \qquad \qquad \qquad \qquad \qquad   \int_{0}^{t} dt \left[m_{11}^{2}(s)  + m_{12 }^{2}(s) \right]ds, \\\\

\mathbb{E} \left\{ U_tV_t\right\} =  \int_{0}^{t}  \left[m_{11}(s) m_{12}(s)  + m_{22 }(s) m_{21}(s)\right]ds, \\\\

\\\\
\mathbb{E} \left\{ V_t^{2}\right\} =  \int_{0}^{t}    \left[m_{22}^{2}(s) + m_{21 }^{2}(s) \right]ds.
\end{array}
\right.
\end{equation}

\noindent Going back to the nominal variables ${\bf z}^{\dagger} = (x,y)$,  Eqns. (\ref{CCO}) and (\ref{OPERATORS}) imply:

\begin{equation}
\label{NOMINAL}
\left\{
\begin{array}{l}
P= \frac{1}{2\pi  \sqrt{{\rm Det}(\mathbb{W}) }} e^{- \frac{{\bf z}^{\dagger} \mathbb{W} (t){\bf z}} {2 {\rm Det}({\mathbb{W}(t)}) }}, \\ \\

\mathbb{W} (t)  = {\cal M}^{\dagger}(t) \Pi(t){\cal M}(t) = \begin{pmatrix}
     A(t) & - H(t)    \\
      -H(t)&   B(t)
\end{pmatrix}, \\\\

\mathbb{E}\left\{X_t^{2} \right\} := \int_{\mathbb{R}} \int_{\mathbb{R}}  x^{2} P\, dxdy =  B(t), \\\\

\mathbb{E}\left\{Y_t^{2} \right\} :=   \int_{\mathbb{R}} \int_{\mathbb{R}}  y^{2} P \, dxdy = A(t),

\end{array}
\right. 
\end{equation}

\noindent with:

\begin{equation}
\label{DETCOR}\left\{ 
\begin{array}{l}
  B(t) = [m_{12}^{2} \Pi_{11}   +m_{12} \Pi_{12} m_{22}  + m_{22} \Pi_{21} m_{12}  + m_{22}^{2} \Pi_{22}  ](t), \\\\

A(t) =  [m_{11}^{2} \Pi_{11}   +m_{11} \Pi_{12} m_{21}  + m_{21} \Pi_{21} m_{11}  + m_{21}^{2} \Pi_{22}  ](t),

\end{array}
\right.
\end{equation}

\noindent where the matrix elements $m_{ij}(t)$ and $\pi_{ij}(t)$ are explicitly  given in Eqns. (\ref{MOINS}) and (\ref{PIPO}). While the present procedure is exact, it leads to cumbersome algebra.

\vspace{0.3cm} 
\noindent {\bf Remark}. Note that a  simpler  case of the above general scheme  is  exposed by S. Chandrasekhar \footnote{S. Chandrasekhar, {\it Stochastic Problems in Physics and Astronomy}. Rev. Mod. Phys,  {\bf 1}, (1943), 1- 84. See Lemma II page 36.} (see Lemma II), for the  simpler case:

$$
\left\{
\begin{array}{l}
{\cal M} = \begin{pmatrix}
     m_{11}(t)  &   m_{11}(t)   \\
      m_{22}(t)  &   m_{22}(t) 
\end{pmatrix}, \\\\
\Delta_{uv} =  m_{11}^{2} (t) \partial_{uu}  +2 m_{11} (t)m_{22}(t) \partial_{uv}+  m_{22}^{2} (t) \partial_{vv}, \\\\
\Pi(t)  = 

\begin{pmatrix}
    2 \int_{0}^{t} m_{22}^{2} (s) ds   &   - \int_{0}^{t}  m_{11}(s)m_{22} (s) ds \\\\
      -   \int_{0}^{t}  m_{11}(s)m_{22} (s) ds  &    2\int_{0}^{t}  m_{11}^{2} (s) ds 
\end{pmatrix}.
\end{array}
\right.
$$

\section{The $(X_t, Y_t)$ stochastic process using coupled dynamics}\label{FULLPRO}

\noindent Consider the stochastic process $(X_t, Y_t) \in \mathbb{R}^{2}$ :

\begin{equation}
\label{SPOINT}
\left\{
\begin{array}{l}

dX_t =\frac{ -\Delta_t  dt+  \sigma_y dW_{1,t} }{t+ \alpha} = \frac{ (y-x)  dt+  \sigma_y dW_{1,t} }{t+ \alpha}, \qquad X_0=x_0

\\\\

dY_t = \frac{ +\Delta_t  dt+  \sigma_x dW_{2,t} }{t+ \beta} = \frac{ (x-y) dt+  \sigma_x dW_{2,t} }{t+ \beta},  \qquad Y_0=y_0
.

\end{array}
\right.
\end{equation}

\noindent  where $dW_{1,t}$ and   $dW_{2,t}$ are independent Gaussian Noise processes (WGN). To study the $(X_t, Y_t)$ bi-variate Gaussian \footnote{The Gaussian property is ensured since the drifts are linear and linear transformations of Gaussian restitutes Gaussian.} and Markovian diffusion process is advantageous to proceed with the change of variables:

\begin{equation}
\label{BB14}
\left\{
\begin{array}{l}
\begin{pmatrix}
      X_t    \\
      Y_t  
\end{pmatrix} \mapsto \begin{pmatrix}
      \Delta_t := X_t- Y_t     \\
      \Sigma_t := X_t+ Y_t 
\end{pmatrix},  \\\\

d \Delta_t = - \frac{[ 2t + \alpha + \beta] }{\left[ (t+ \alpha )(t+ \beta) \right] } \Delta_t dt+

 \frac{\sqrt{\left[ \sigma_{x}^{2} (t+ \beta)^{2} + \sigma_y^{2} (t+ \alpha )^{2}\right]} d B_{1,t}}{\left[ (t+ \alpha )(t+ \beta) \right] }, \\\\ 

  d\Sigma_t = +\frac{ (\alpha - \beta )}{\left[ (t+ \alpha )(t+ \beta) \right] }\Delta_t dt  + \, \frac{\sqrt{\left[ \sigma_{x}^{2} (t+ \beta)^{2} + \sigma_y^{2} (t+ \alpha )^{2}\right]}
  dB_{2,t}}{\left[ (t+ \alpha )(t+ \beta) \right] }, 
  \end{array}
\right.
\end{equation}

\noindent where we have used the property:

\begin{equation}
\label{NSOURCES}
\left\{
\begin{array}{l}
\frac{\sigma_y}{(t+\alpha) } dW_{1,t} - \frac{\sigma_x}{(t+\alpha) } dW_{2,t}  =   
 \frac{\sqrt{\left[ \sigma_{x}^{2} (t+ \beta)^{2} + \sigma_y^{2} (t+ \alpha )^{2}\right]} d B_{1,t}}{\left[ (t+ \alpha )(t+ \beta) \right] },
\\\\

\frac{\sigma_y}{(t+\alpha) } dW_{1,t} +  \frac{\sigma_x}{(t+\alpha) } dW_{2,t}  = 
 \frac{\sqrt{\left[ \sigma_{x}^{2} (t+ \beta)^{2} + \sigma_y^{2} (t+ \alpha )^{2}\right]} d B_{2,t}}{\left[ (t+ \alpha )(t+ \beta) \right] }
 \end{array}
\right.
\end{equation}

 \noindent with $dB_{1,t}$ and $dB_{2,t}$  being now  {\bf correlated}  WGN's. Since the $\Delta_t$ process is actually decoupled from the $\Sigma_t$, we shall proceed in two steps.
 
\subsubsection{The $\Delta_t$ process}\label{DELTATIME}
 \noindent  The  probabilistic properties of $\Delta_t$ stochastic process in Eqn. (\ref{BB14}) are fully described by the TPD $P_{\Delta}(x,t |x_0,0)dx:= {\rm Prob} \left\{x \leq \Delta_t\leq(x+dx)|x_0 \right\} $  which solves the FPE:

\begin{equation}
\label{FPFULL}
\left\{
\begin{array}{l}

  \partial_tP_{\Delta} =  \partial_x \left\{ \left[  \frac{[ 2t + \alpha + \beta] }{\left[ (t+ \alpha )(t+ \beta) \right] } \right] x\right\}  + D(t) \partial_{xx} P_{\Delta}, \\\\
D(t) := \frac{\sigma_y^{2}(t+ \beta)^{2} + \sigma_x^{2}(t+ \alpha)^{2}}{2(t+ \alpha)^{2} (t+ \beta)^{2}}.
\end{array}
\right.
\end{equation}

\noindent To solve Eqn.(\ref{FPFULL}), similarly to Appendix \ref{LIOUVILLE}, we express the evolution in terms of the constant  of the motion $\Delta_o$. Accordingly,  we introduce the change of variables:

\begin{equation}
\label{CV1}
\left\{
\begin{array}{l}
 t \mapsto \tau = t  \,\,\, \ \Rightarrow \quad \partial_t \mapsto \frac{\partial_{x'} }{\partial_t}\partial_{x'} + \frac{\partial \tau }{\partial_t} \partial_{\tau} =\left[ \frac{2t + \alpha + \beta}{\alpha\beta} \right] x\partial_{x'}+  \partial_{\tau}  = \\\\
 
 \qquad  \qquad  \qquad  \qquad  \qquad  \qquad  \qquad  \qquad  \qquad  \qquad  \qquad  \qquad  \left[ \frac{2t + \alpha + \beta}{(t+ \alpha )(t+\beta) } \right]  x'\partial_{x'} + \partial_{t},
  \\\\
 x \mapsto x' :=  x\frac{(t+ \alpha) (t + \beta)}{\alpha \beta} \quad \Rightarrow \quad \partial_x \mapsto  
 \frac{\partial x'}{\partial_x } \partial_{x'} + \frac{\partial \tau}{\partial x} \partial_{\tau} = \frac{(t+\alpha) (t+ \beta)}{\alpha \beta} \partial_{x'}
\end{array}
\right.
\end{equation}

\noindent In terms  of the $(t,x')$, Eqn. (\ref{FPFULL}) is transformed into a pure diffusion process. This can be seen as follows, (omitting the arguments of $P$):

$$
\begin{array}{l}
 \partial_{t} P_{\Delta}  +\left[ \frac{2t + \alpha + \beta}{(t+ \alpha)(t +\beta) } \right] x' \partial_{x'} P_{\Delta} =   \\\\ 
 
 \qquad \quad 
  \frac{(\alpha + t) (\beta + t)}{\alpha \beta}  \partial_{x'}  \left\{  \frac{[ 2t + \alpha + \beta] }{\left[ (t+ \alpha )(t+ \beta) \right] } \frac{\alpha\beta} {(t + \alpha) (t + \beta)}  x'P \right\} +  D(t)  \frac{(\alpha+ t)^{2} (\beta+ t)^{2}}{\alpha^{2} \beta^{2}} \partial_{x' x'}P_{\Delta} = \\\\ 
   \qquad \qquad  \qquad \qquad  \qquad \quad  \qquad \quad 
   \frac{[ 2t + \alpha + \beta] }{\left[ (t+ \alpha)(t+ \beta) \right] }  \partial_{x'} (x'P_{\Delta}) + 
   D(t)  \frac{(\alpha + t)^{2} (\beta + t )^{2}}{\alpha^{2} \beta^{2}} \partial_{x'x'}P_{\Delta},
 
 \end{array}
$$

\noindent yielding :

\begin{equation}
\label{FPFULL2}
\begin{array}{l}
 \partial_{t }P_{\Delta}   =    \frac{[ 2t+ \alpha + \beta] }{\left[ (t+ \alpha )(t+ \beta) \right] } P_{\Delta}+ 
  D(t) \frac{(\alpha + t)^{2} (\beta + t)^{2}}{\alpha^{2} \beta^{2}} \partial_{x'x'}P_{\Delta}.
 
 \end{array}
\end{equation}

\noindent Writing $P_{\Delta} := (t + \alpha) (t + \beta) Q$, 
%
%
Eq(\ref{FPFULL2}), reduces to the  pure time  inhomogeneous  diffusion:

\begin{equation}
\label{FPFULL3}
\partial_{t} Q_{\Delta} =\left[  D(t)\frac{(\alpha + t) (\beta + t)}{\alpha^{2} \beta^{2}} \right]\partial_{x'x'}Q_{\Delta} =\left[ \frac{\sigma^{2}_x(t + \alpha)^{2} + \sigma^{2}_y(t + \beta)^{2} }{2 \alpha^{2} \beta^{2}(\alpha+t)(\beta+t)} \right]
\partial_{x'x'}Q_{\Delta},
\end{equation}


\noindent Finally,  we  introduce the time re-scaling :

\begin{equation}
\label{RESCAL}
t  \mapsto  s(t): = \int_{0}^{t} \left[ \frac{\sigma^{2}_x( \alpha + \xi)^{2} + \sigma_y^{2}(\beta + \xi )^{2} }{\alpha^{2} \beta^{2}(\alpha+\xi)(\beta+\xi)} \right]d\xi = \frac{(\sigma_x^2+\sigma_y^2)t+(\alpha-\beta)(\sigma_x^2\log\tfrac{\beta+t}{\beta}-\sigma_y^2\log\tfrac{\alpha+t}{\alpha})}{\alpha^2\beta^2},
\end{equation}

\noindent This enables to rewrite Eqn. (\ref{FPFULL3}) as:

\begin{equation}
\label{BM}
 \partial_s Q_{\Delta} = \frac{1}{2} \partial_{yy} Q_{\Delta} \qquad \Rightarrow \qquad  Q_{\Delta} = \frac{e^{- \frac{(y-y_0)^{2}}{2s(\tau)}}}{\sqrt{2 \pi s(\tau)} }.
\end{equation}

\noindent Proceeding backwards  to the nominal  $(x,t)$ variables,  one ends with:

\begin{equation}
\label{FINO}
\begin{array}{l}
P_{\Delta}(x, t|x_0,0) dx =(t + \alpha) (t + \beta)  \frac{e^{ -\frac{ \left[x\frac{(t+ \alpha) (t+ \beta)}{\alpha \beta}-x_0)\right]^{2}}{2s(\tau)}}}{  \alpha \beta\sqrt{2 \pi s(t)} } dx =  \\\\

\qquad \qquad \qquad \qquad \qquad \qquad \qquad 
\frac{1}{\sqrt{2 \pi \sigma^{2}_{\Delta}(t)}}  e^{- \frac{\left[x- \frac{\alpha\beta \, x_0}{(t + \alpha) (t+ \beta)}  \right]^{2} }{2 \sigma^{2}_{\Delta}(t)} }, 
\end{array}
\end{equation}

\noindent where we used the notation:

\begin{equation}
\label{SD}
\left\{
\begin{array}{l}

\sigma_{\Delta}^{2} (t): = \frac{\alpha^{2} \beta^{2} s(t) } {(t+ \alpha)^{2} (t+ \beta)^{2} }  =  
\frac{(\sigma_x^2+\sigma_y^2)t+(\alpha-\beta)\big(\sigma_x^2\log\tfrac{\beta+t}{\beta}-\sigma_y^2\log\tfrac{\alpha+t}{\alpha}\big)}{ (t+ \alpha)^{2} (t+ \beta)^{2}  }, \\\\
 \displaystyle \lim_{t \rightarrow \infty} \sigma_{\Delta}^{2} (t)=0.
 \end{array}
 \right.
\end{equation}

\subsection{Non-monotonous relaxation of the variance $ \sigma_{\Delta}^{2} (t)$} \label{relaxationtime}

\noindent As $X_0=x_0$ and $Y_0=y_0$ are fixed and deterministic, we obviously have  $\displaystyle \lim_{t \rightarrow \infty} \sigma_{\Delta}^{2} (t)=0$, In parallel, from Eqn. (\ref{SD})   we have $\displaystyle\lim_{t\rightarrow \infty} \Delta_t =0$.  Since  $\sigma_{\Delta}^{2} (t)\geq 0$, we conclude   that $\sigma_{\Delta}^{2} (t)$  follows a non-monotonous evolution reaching a maximum at a relaxation time $t^{*}$ such that $\frac{d}{dt } \sigma_{\Delta}^{2} (t) \mid_{t= t^{*}}=0$. On physical grounds, this non-monotonous evolution describes the underlying  trade-off between  two distinct mechanisms:  a  disorganising mechanism generated  by the  noisy driving forces versus  the   organising  mechanism generated by the  mutual interactions.  During  the early stage  $0< t< t^{*}$, the fluctuations dominate while later for  $t>t_c$ the learning mechanism  overcomes the underlying  noise  to  ultimately  drives $\sigma_{\Delta}(t)$ towards zero. Accordingly, it is legitimate to interpret $t_c$ as a relaxation time.

 
\subsection{The $\Sigma_t$ process - approximation for the time asymptotic regime}\label{SIGMATIME}

\noindent  Since the full transient evolution of variances as given in Appendix \ref{LIOUVILLE} leads to cumbersome expressions, 
let us focus on the time asymptotic development. We already know  exactly that $\displaystyle  \lim_{t \rightarrow \infty} \mathbb{E}\left\{\Delta_t\right\} =  \lim_{t \rightarrow \infty}  \mu_{\Delta}(t)=0$,  and from the last section we have  $\displaystyle \lim_{t \rightarrow \infty} \sigma_{\Delta}(t) =0$. Accordingly, in the time asymptotic regime, the initially bi-variate diffusion collapses to  a scalar (i.e., uni-variate) pure diffusion  $\Sigma_t$  process centered at the constant final value $\mu_{\Sigma, f}$ given in Eq.(\ref{LIMO}. Hence for asymptotic times  Eqn. (\ref{BB14}), the  $\Sigma_t$  evolution can be approximately written as:

\begin{equation}
\label{SIGINF}
d(\Sigma_t - \mu_{\Sigma, f} ) = d\Sigma_t =  \frac{\sqrt{\left[ \sigma_{x}^{2} (t+ \beta)^{2} + \sigma_y^{2} (t+ \alpha )^{2}\right]}
  }{\left[ (t+ \alpha )(t+ \beta) \right] }dB_{2,t}.
\end{equation}

\noindent The  associated TPD  $ P_{\Sigma}(x,t|\Sigma_0,0)dx= {\rm Prob}\left\{x \leq \Sigma_t \leq (x+dx)|\Sigma_0 \right\} dx:= P_{\Sigma}dx$ obeys to  the FPE:

\begin{equation}
\label{FPSIGINF}
\partial_t P_{\Sigma}  =  \frac{\left[ \sigma_{x}^{2} (t+ \beta)^{2} + \sigma_y^{2} (t+ \alpha )^{2}\right]}
  {(t+ \alpha )^{2} (t+ \beta)^{2} } \frac{1}{2}\partial_{xx}P_{\Sigma}.
\end{equation}

\noindent To solve  Eqn. (\ref{FPSIGINF}), as usual we introduce the re-scaling:

\begin{equation}
\label{TREASC}
t \mapsto s(t) := \int_{0}^{t}\frac{\left[ \sigma_{x}^{2} (\xi+ \beta)^{2} + \sigma_y^{2} (\xi+ \alpha )^{2}\right]}
  {(\xi+ \alpha )^{2} (\xi+ \beta)^{2} } d\xi = \sigma_x^{2} \left[ \frac{1}{\alpha} - \frac{1}{\alpha+ t}\right]  +  \sigma_y^{2} \left[ \frac{1}{\beta} - \frac{1}{\beta+ t}\right],
\end{equation}

\noindent yielding:

\begin{equation}
\label{PDEL}
\left\{
\begin{array}{l}
 P_{\Sigma}(x,t|\Sigma_0,0)= \frac{1}{\sqrt{2 \pi s(t)}} e^{- \frac{[x-\mu_{\Sigma, f} ]^{2}}{2 s(t)} }, \\\\

 \lim_{t \rightarrow \infty}   P_{\Sigma}(x,t|\mu_{\Sigma, f},0) = \frac{1}{\sqrt{2 \pi s_{\infty}}} e^{- \frac{[x-\mu_{\Sigma, f}]^{2}}{2 s_{\infty}} }, \qquad s_{\infty}=\left[ \frac{\sigma^{2}_{x}}{\alpha} + \frac{\sigma^{2}_{y}}{\beta} \right], \\\\
 
 \displaystyle \lim_{t \rightarrow \infty}  \sigma^{2}_{\Sigma}(t)  = \mathbb{E} \left\{ (\Sigma_t - \mu_{\Sigma, f} )^{2}\right\}   = s_{\infty}.
 \end{array}
 \right.
\end{equation}

\noindent Finally, for asymptotic times, we have that $ \displaystyle \lim_{t \rightarrow \infty} X_t = \lim_{t \rightarrow \infty} Y_t $, we can conclude that the asymptotic variances of the $X_t$ and $Y_t$  converge to:

\begin{equation}
\label{ASYMPXY}
 \displaystyle \lim_{t \rightarrow \infty}  \sigma^{2}_{X}(t) =  \displaystyle \lim_{t \rightarrow \infty}  \sigma^{2}_{Y} (t).   \\\\
\end{equation}

\end{widetext}

\bibliographystyle{ieeetr}
\footnotesize\bibliography{bb}
\end{document}